\documentclass[aps,prb,showpacs,twocolumn,amsmath,amssymb]{revtex4}
\usepackage{graphicx}
\begin{document}
\title{High frequency dynamics in liquid nickel: an IXS study}
\author{S. Cazzato $^{1,2}$, T. Scopigno $^{1,2}$, S. Hosokawa $^3$, M. Inui $^4$, W-C. Pilgrim $^5$ and G. Ruocco$^{1,2}$}
\affiliation{$ˆ1$ Dipartimento di Fisica and INFM, Universit\`{a}
di Roma "La Sapienza", I-00185 Roma, Italy. } \affiliation{$ˆ2$
INFM CRS-SOFT, c/o Universit\`{a} di Roma "La Sapienza", I-00185,
Roma, Italy.} \affiliation{$ˆ3$ Department of Materials Science,
Hiroshima University, Higashi-Hiroshima 739, Japan.}
\affiliation{$ˆ4$ Faculty of Integrated Arts and Sciences,
Hiroshima University, Higashi-Hiroshima 739, Japan.}
\affiliation{$ˆ5$ Institute of Physical Chemistry and Materials Science Centre, Philipps-University of Marburg, 35032 Marburg, Germany.} 
\begin{abstract}
Owing to their large relatively thermal conductivity, peculiar,
non-hydrodynamic features are expected to characterize the
acoustic-like excitations observed in liquid metals. We report
here an experimental study of collective modes in molten nickel, a
case of exceptional geophysical interest for its relevance in
Earth interior science. Our result shed light on previously
reported contrasting evidences: in the explored energy-momentum
region no deviation from the generalized hydrodynamic picture
describing non conductive fluids are observed. Implications for
high frequency transport properties in metallic fluids are
discussed.
\end{abstract}

\maketitle
\section{INTRODUCTION}\label{sec:Introduction}
Half a century of inelastic neutron scattering experiments,
recently complemented by similar investigations with X-rays, have
clearly shown that acoustic like excitations can be sustained by
simple liquids down to wavelengths comparable to the mean
interparticle distances and frequencies extending up to the THz
region.\cite{HANSEN,BALUCANI,scop_rmp} In ordinary, non conductive
fluids, such collective modes can be roughly described in terms of
adiabatic sound waves of energy $E=\hbar \Omega$ propagating with
sound velocity $c_s=\lim_{Q\rightarrow 0}\frac{\Omega (Q)}{Q}$.
This is a well known, immediate consequence of the hydrodynamic
treatment when the condition $\Omega>>D_T Q^2$ holds, where $D_T$
is the thermal diffusion coefficient for the system under study
and $Q$ the exchanged momentum between the probe and the
sample.\cite{HANSEN,scop_rmp} A deeper analysis reveals that such
acoustic waves are actually subject to relaxation processes
related to the frequency dependent viscosity. The origin of such
processes is twofold: the structural relaxation, responsible for
dynamical arrest in those systems capable of supercooling, and a
microscopic process which induces an additional damping of the
sound waves due to the non-plane wave nature of the instantaneous
vibrational eigenmodes. Both these processes manifest themselves
at wavevectors $Q\approx 1/10 Q_M$, being $Q_M$ the principal
maximum of the static structure factor. This region,
unfortunately, represents the technical limit for both neutron and
X-ray inelastic spectroscopy, hindering the complete understanding
of the underlying dynamical processes which are still
controversially debated. \cite{scop_rmp}

In liquid metals, an additional complication arises. The
hydrodynamic condition $\Omega (Q)>>D_T Q^2$ breaks down in the
range $0.1\lesssim Q \lesssim 1$ nm$^{-1}$, an estimate obtained
neglecting the $Q$ dependence of the thermal conductivity.
Consequently, an isothermal regime may be expected to occur above
the $\Omega (Q) \approx D_T Q^2$ crossover, with the adiabatic
limit attained only below this value. \cite{scop_rmp,FABER,
Scop_liq07} Among the implications of such an isothermal regime on
the sound waves propagation, noteworthy features would be:
\begin{itemize}
\item a reduced value of the sound velocity
$c_t=c_s/\sqrt{\gamma}$, with $\gamma=C_P/C_V$, the ratio of
constant pressure to constant volume specific heats. \item a
different expression for the thermal contribution to the acoustic
attenuation: $\frac{(\gamma-1)c_t^2}{\gamma D_T}$ instead of
$(\gamma -1) D_T Q^2$.
\end{itemize}
To address this issue, we present here an investigation of the
high frequency dynamics in liquid nickel performed by Inelastic
X-ray Scattering (IXS), a technique which has proved, since the
early 90's, to be an invaluable tool for deepening our
comprehension of the dynamics of simple liquids on the microscopic
scale, since it allows on one hand to overcome the kinematic
limitations due to the lower incident energies of the neutrons and
their specific energy-momentum relation, and on the other hand
provides direct access to the coherent cross section of the
scattering process. \cite{scop_rmp}

In this respect, previous investigations performed by means of
Inelastic Neutron Scattering (INS) and molecular dynamics (MD)
have shown contrasting results. \cite{ber_ni,alem_sim, alem_trans,
cher} In the former study the sound velocity attains indeed the
isothermal value at the longest accessible wavelength ($Q\approx
8\mathrm{\:nm^{-1}}$), while in the latter the numerically
estimated sound velocity is always larger than the adiabatic
value. Recently an interesting study involving both quasi-elastic
Neutron Scattering and MD has pointed out how the simulation data
for the speed of sound of liquid Ni attain the adiabatic value
$c_S$ at the lower accessed $Q$'s ($Q\approx 2
\mathrm{\:nm^{-1}}$). \cite{ruiz}

Liquid Ni, indeed, is characterized by the largest specific heat
ratio among monatomic liquids, a property which should emphasize
any non hydrodynamic results. In addition, Ni is a system of
paramount relevance in geophysical science due to its presence in
the Earth's interior. \cite{poi_earth} Indeed, the fact that the
Earth outer core is mostly iron was established beyond reasonable
doubt already in the early 60's, when it was confirmed that the
density of the core was about 10\% lower than the density of iron,
and that the seismic parameter $\Phi =K / \rho$, being $K$ the
bulk modulus and $\rho$ the density, was higher than that of iron.
\cite{Birch_Geo_64} About 4\% Ni is thought to be present in the
core and although it does not appreciably change the density of
liquid Fe, its presence should not be forgotten as phase diagrams
of systems Fe-Ni-light elements may be significantly different
from those of systems without Ni. \cite{Bret_Geo_71, Wald_Geo_04}
Among the most favorite candidates as lighter alloying elements
with Fe and Ni, sulfur is perhaps the most addressed one.
\cite{poi_earth} In this respect, recently a high anomalous
behavior of the ultrasonic sound velocity $c_S(T)$ and attenuation
$\Gamma(T)$ as functions of temperature was reported in the
mixture 85\%Fe-5\%Ni-10\%S, for temperatures above melting
($\mathrm{T_m}=$1650 K) up to 2000 K, at ambient pressure
conditions.\cite{Nash_FeNiS} In fact, and contrary to the data of
pure liquid metal components, the acoustic velocity is found to
increase with temperature, as well as attenuation. A complete
understanding of the underlying behavior of the alloy requires, in
our opinion, to ascertain the nature of acoustic excitations in
pure liquid Ni.

\section{THEORETICAL BACKGROUND: EXPECTED HYDRODYNAMIC BEHAVIOR FOR LIQUID METALS}
 In an IXS experiment the double differential cross-section,
which depends on the exchanged momentum $Q$ and energy $E$, is
proportional to the so called dynamic structure factor
$S(Q,\omega)$, which in turn is the Fourier transform of the time
dependent intermediate scattering function
\begin{equation}
F(Q,t)=\frac{1}{N}\sum_{i,j}\left< e^{-iQ\cdot r_i(0)}e^{iQ\cdot
r_j(t)}\right>\text{.} \label{eq:intermediatetime}
\end{equation}
Here $N$ is the total number of particles constituting the system,
and $r_i(t)$ the position of particle $i$ at time $t$.
\cite{scop_rmp} In particular, the zero time value of the
intermediate scattering function is directly related to the
structural features of the system, being $F(Q,0)=S(Q)$, i.e. the
static structure factor. In the hydrodynamic limit, for low values
of $Q$, the dynamic structure factor displays three distinct
peaks. A quasi elastic one, located at zero energy exchange, whose
width $\Gamma_e=D_TQ^2$ is related to the thermal diffusion
coefficient $D_T$, and two inelastic peaks - the so called
Brillouin doublet - located at frequencies $\omega=\pm c_SQ$,
\footnote{Up to first order in $Q$, but this relation becomes
exact if we define the sound velocity from the maxima of the
current $J(Q,\omega)=\omega^2/Q^2S(Q,\omega)\;$.} with $c_S$ being
the adiabatic speed of sound, and whose width depends mostly on
kinematic viscosity, which is the main mechanism driving sound
damping. On the basis of linear hydrodynamics, the dispersive
behavior, i.e. the dependence of the frequency $\omega$ of
propagating collective modes on $Q$, is well known to display a
transition between a linear adiabatic regime to a linear
isothermal one, characterized by an isothermal speed of sound
$c_T$, at $Q$ values such that $\omega \tau_{th}\sim 1$, where
$\tau_{th}$ is the characteristic decay time of thermal
fluctuations ($\tau_{th}\sim1/D_TQ^2$).\cite{scop_rmp,FABER} In
other words, at sufficiently high $Q$ values, thermal fluctuations
are expected to decay on a timescale much longer than the
timescale of sound propagation, which now takes place in a
thermalized environment. If we look at the lineshape of the
dynamic structure factor, the linewidth of the quasi-elastic line
will increase, ultimately overwhelming the Brillouin doublet and
causing a shift of the inelastic peaks position. The adiabatic and
isothermal speed of sound, as we already pointed out in section
\ref{sec:Introduction}, are related by the specific heat ratio
$\gamma$, and we may call $Q_0\sim c_S/D_T$ the value at which the
adiabatic to isothermal transition takes place. In this scenario
liquid metals constitute a particularly interesting class of
simple liquids, because of their high thermal diffusivity with
values of about ten times larger than $c_S$, shifting the
$Q$-range of the isothermal region down to values between $0.1
\mathrm{\:nm^{-1}}$ and $3 \mathrm{\:nm^{-1}}$ provided that:
\begin{enumerate}
    \item the $Q$ dependence of transport coefficients can be neglected
    below at least $3\mathrm{\:nm^{-1}}$, as is the case
    encountered in most liquid metals.
    \item the system interaction with radiation can be described
    within an effective single-component model, in which only the core
    electrons from the metallic ions interact with the
    electromagnetic radiation. However transport coefficients are
    supposed to take an effective value also carrying the net
    effect due to valence electrons.
\end{enumerate}
\begin{table}
\caption{Thermal properties of selected liquid metals near the
melting point. $\gamma$, the ratio of specific heats, and $D_T$,
the thermal diffusion coefficient, are reported at those
temperatures T, near the melting point, where data are available.
For a more extensive summary of data on thermal as well as
dynamical properties of liquid metals, see Ref. 3.
\label{tab:ThermalProperties} }
\begin{center}
\begin{tabular}{llcc}
\hline
Sample&$T [\mathrm{K}]$&$\gamma$&$D_T [\mathrm{nm^2/ps}]$\\
\hline \\
$\mathrm{Li}$&453&1.08$^{\cite{ida,hult}}$1.065$^{\cite{OSE}}$&19.1$^{\cite{tpm_DT}}$,20.3$^{\cite{tpm_DT}}$\\
$\mathrm{Na}$&371&1.12$^{\cite{klep}}$1.091$^{\cite{OSE}}$&68.8$^{\cite{tpm_DT}}$\\
$\mathrm{Mg}$&923&1.29$^{\cite{ida}}$&39.8$^{\cite{tpm_DT}}$\\
$\mathrm{Al}$&933&1.4$^{\cite{ida}}$&35.2$^{\cite{tpm_DT}}$\\
$\mathrm{K}$&336.7&1.11$^{\cite{klep}}$1.102$^{\cite{OSE}}$&81.4$^{\cite{tpm_DT}}$\\
$\mathrm{Fe}$&1808&1.8$^{\cite{ida}}$&7.3\cite{tpm_DT}\\
$\mathrm{Co}$&1765&1.8$^{\cite{ida}}$\\
$\mathrm{Ni}$&1728&1.98$^{\cite{ida}}$&9.6${\cite{ida}}$\\
&1763&1.88$^{\cite{ber_ni}}$\\
$\mathrm{Cu}$&1356&1.33$^{\cite{ida}}$&42.1$^{\cite{tpm_DT}}$\\
$\mathrm{Zn}$&693&1.25$^{\cite{klep}}$1.26$^{\cite{ida}}$&15.7$^{\cite{tpm_DT}}$\\
$\mathrm{Ga}$&303&1.08\cite{klep}&11.6\cite{tpm_DT}\\
$\mathrm{Ge}$&1253&1.18$^{\cite{ida,hos_ge}}$\\
$\mathrm{Rb}$&312&1.15$^{\cite{klep}}$,1.097$^{\cite{OSE}}$&61.5$^{\cite{tpm_DT}}$\\
$\mathrm{Ag}$&1233&1.32$^{\cite{ida}}$&66.5$^{\cite{tpm_DT}}$\\
$\mathrm{Sn}$&505&1.11$^{\cite{klep}}$&17.3$^{\cite{tpm_DT}}$\\
$\mathrm{Cs}$&302&&44.6$^{\cite{tpm_DT}}$\\
&308&1.102$^{\cite{bod_cs}},$&\\
$\mathrm{Au}$&1336&1.28$^{\cite{ida}}$&40.4$^{\cite{tpm_DT}}$\\
$\mathrm{Hg}$&293&1.14$^{\cite{bad_hg}}$&4.41$^{\cite{tpm_DT}}$\\
$\mathrm{Pb}$&623&1.19$^{\cite{ida,hult}}$&9.89$^{\cite{tpm_DT}}$\\
\hline
\end{tabular}
\end{center}
\end{table}
 The second assumption has been on the basis of a
recent debate focused on the interpretation of IXS data on liquid
alkali metals and aluminum. \cite{sing_pre, scop_comm, sing_rep}
The main difficulty in ascertaining the existence of sound
propagation with lower than adiabatic sound velocity value lies in
the fact that $\gamma$ is for most metallic systems very close to
unity, as can be seen from the specific heats ratios reported for
a selection of liquid metals near their melting point in table
\ref{tab:ThermalProperties}. For alkali metals $\gamma\sim1$ as
well as for the great majority of liquid metals. Among the systems
exhibiting a comparably high value of the specific heat ratio (Fe,
Co and Ni), liquid nickel has been chosen, also in view of its
lower - despite still very high - melting temperature.

\section{The experiment}
The experiment reported in this work was carried out at the high
resolution Beamline ID16 of the European Synchrotron Radiation
Facility (Grenoble, Fr).
\begin{figure}[h]
\includegraphics[width=0.47\textwidth]{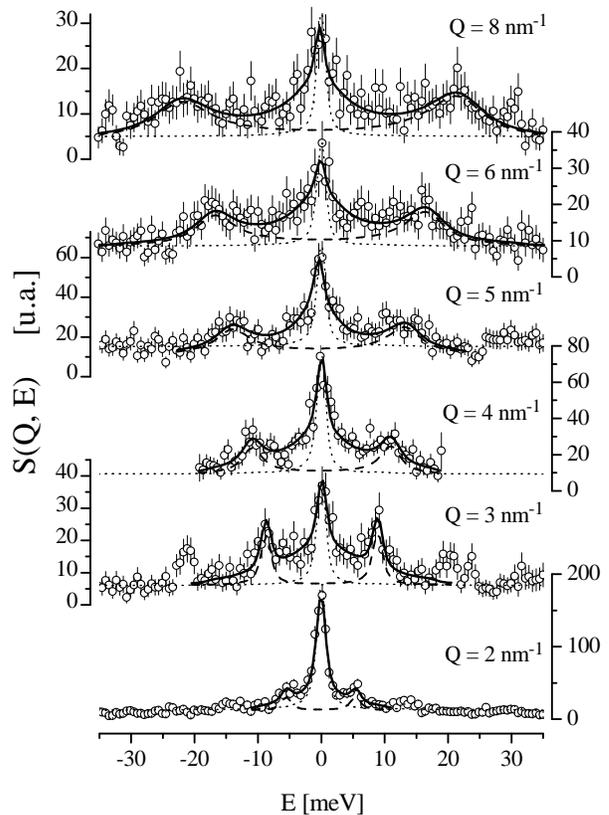}
\vspace{-1.2cm} \caption{(Color online) $\mathbf{a}$: sound
dispersion for liquid Ni: IXS ($\square$) from the present work;
an INS investigation by Bermejo {\it et. al.} \cite{ber_ni}
($\blacktriangle$); a MD simulation from Ruiz-Mart\'{i}n {\it et
al.} ($-\cdot-$) \cite{ruiz}. Adiabatic (---) and isothermal
($---$) sound dispersions are also reported. $\mathbf{b}$: Speed
of sound reported from the present work ($\square$) and from INS
\cite{ber_ni} ($\blacktriangle$) and MD ($-\cdot-$ \cite{ruiz} and
$\circ$ \cite{alem_sim}) investigations. The adiabatic, $c_S$, and
isothermal, $c_T$, speed of sound are displayed
($\cdot\cdot\cdot$).}\label{fig:NiSpectra}
\end{figure}
The backscattering monochromator and analyzer crystals, operating
at the $(11,11,11)$ silicon reflections gave a total energy
resolution of 1.5 meV, while energy scans were performed by
varying the temperature of the monochromator with respect to that
of the analyzer crystals. The sample, in the shape of a
100$\mathrm{\: \mu m}$ thick foil, was 99.993\% purity nickel,
purchased by Rare Metallic Co. Ltd. (Japan), and was hold in a
sapphire cell obtained from a single monocrystal. \cite{tam_cell1}
The use of sapphire prevented the sample interaction with
container, still providing a good cell transmission, with a total
cell thickness of 500 $\mathrm{\mu m}$ seen by the scattered beam.
The cell was lodged in a molybdenum holder, and heated up by means
of a properly isolated tungsten resistance. The nickel absorption
length at an energy of the incoming beam of 21 KeV, is about
50$\mathrm{\: \mu m}$, thus in our condition a 15\% sample
transmission was expected. A selection of IXS spectra from liquid
Ni at 1767 K is reported in Fig. \ref{fig:NiSpectra} as open
circles at the lowest accessed $Q$ values. Dotted lines report the
experimental resolution. Phonon modes from the sapphire cell
(speed of sound of about $11\mathrm{\:kms^{-1}}$) are well
recognizable up to 3$\mathrm{\:nm^{-1}}$, however they are well
separated from inelastic features from the sample at $Q$ greater
than 1$\mathrm{\:nm^{-1}}$.

\section{RESULTS AND DISCUSSION}
 Experimental data have been analyzed with a damped
harmonic oscillator (DHO) function centered at frequency
$\omega_L$, and of width $\Gamma_L$ modelling inelastic
contributions from the metal. The elastic features have been
represented by a Lorentzian of full width at half maximum (FWHM)
$\Gamma_C$. Such an approximation is expected to work well for not
too high $Q$'s, i.e. until there is a clear separation between
elastic and inelastic features, as is the present case for IXS
measurements.
\begin{figure}[ht]
\centering
\includegraphics[width=0.55\textwidth]{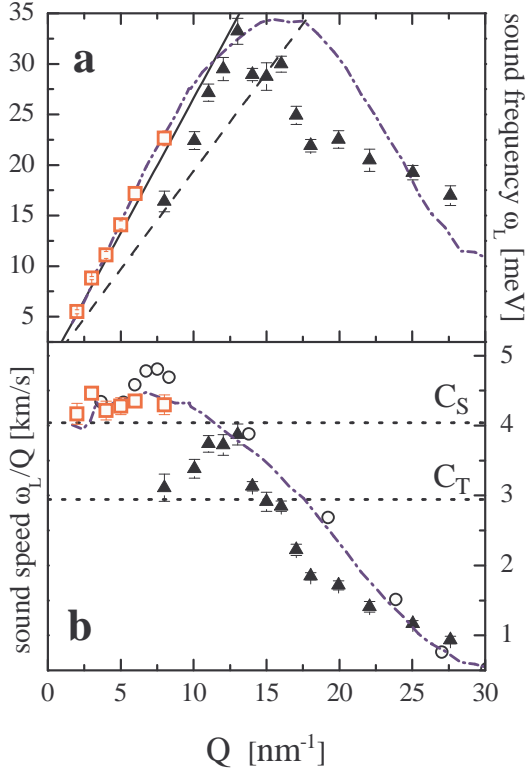}
\vspace{-3.4cm} \caption{\label{fig:speedNi} (Color online)
$\mathbf{a}$: sound dispersion for liquid Ni: IXS ($\square$) from
the present work; an INS investigation by Bermejo {\it et. al.}
\cite{ber_ni} ($\blacktriangle$); a MD simulation from
Ruiz-Mart\'{i}n {\it et al.} ($-\cdot-$) \cite{ruiz}. Adiabatic
(---) and isothermal ($---$) sound dispersions are also reported.
$\mathbf{b}$: Speed of sound reported from the present work
($\square$) and from INS \cite{ber_ni} ($\blacktriangle$) and MD
($-\cdot-$ \cite{ruiz} and $\circ$ \cite{alem_sim})
investigations. The adiabatic, $c_S$, and isothermal, $c_T$, speed
of sound are displayed ($\cdot\cdot\cdot$).}
\end{figure}
\begin{figure}[ht]
\centering
\includegraphics[width=0.543\textwidth]{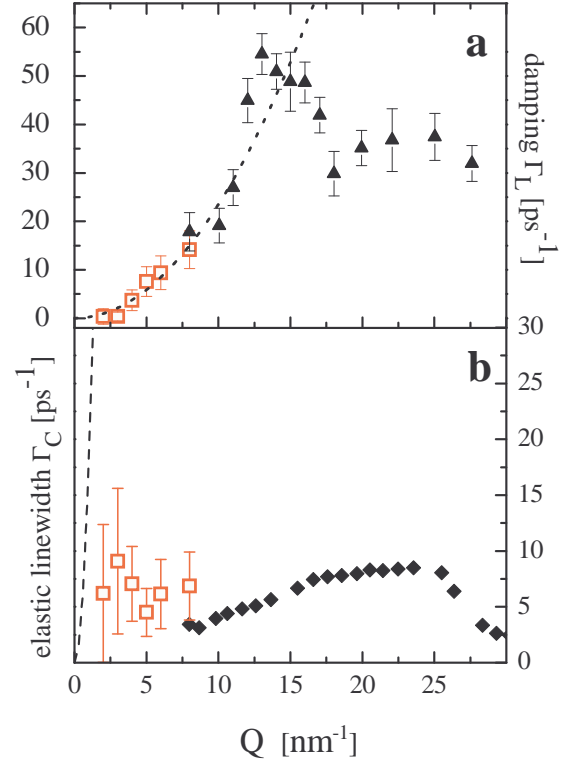}
\vspace{-3.26cm} \caption{(Color online) $\mathbf{a}$: sound
damping from the present work ($\square$) and from INS
\cite{ber_ni} ($\blacktriangle$). The dashed line ($---$) is a
quadratic best fit to the IXS data. $\mathbf{b}$: the quasielastic
linewidth obtained in the present investigation of liquid Ni
($\square$) and from a recent INS investigation (from the coherent
dynamic structure factor) \cite{ruiz} ($\blacklozenge$). The
expected hydrodynamic behavior is also reported ($---$), from the
knowledge of the thermal diffusion coefficient $D_T$.
\cite{ida}}\label{fig:dampNi}
\end{figure}
At $Q$ values approaching $Q_M$, corresponding to the static
structure factor maximum, one has to resort to a model based on a
so-called extended heat mode and two extended sound modes, also
incorporating the frequency sum rules. \cite{Mont_97} The elastic
contribution from the cell, supposed to be much narrower than the
experimental resolution, has been modelled with a Dirac delta
function centered at zero frequency. Thus the model for the
$S(Q,\omega)$ can be written as
\begin{eqnarray}
I_q(\omega)=I_{C,1}(Q)\delta(\omega)+I_{C,2}(Q)\frac{\Gamma_C(Q)}{\omega^2
+\Gamma_C(Q)^2}+\nonumber\\
+I_L(Q)\frac{\omega_L^2(Q)\Gamma_L^2(Q)}{\left(\omega^2-\omega_L^2(Q)\right)^2
+\left(\omega\Gamma_L(Q)\right)^2}\text{.} \label{eq:model}
\end{eqnarray}
Additionally, in order to properly reproduce the experimental
spectra, the above expression has to be modified such as to be
compliant with the detailed balance condition:
\begin{equation}
S(Q,\omega)=\frac{\beta\hbar\omega}{1-e^{-\beta\hbar\omega}}I_q(\omega)\text{.}
\label{eq:detailedBalance}
\end{equation}

Before comparison with IXS data, the model is convoluted with the
experimental resolution. This procedure results in the curves
displayed as full lines in Fig. \ref{fig:NiSpectra}. Fig.
\ref{fig:speedNi} reports the dispersion (a) and sound velocity
(b) obtained for liquid Ni from the DHO model parameter $\omega_L$
at different $Q$ values ($\square$), while Fig. \ref{fig:dampNi}.a
reports sound damping as derived from the DHO parameter
$\Gamma_{L}$ ($\square$). Sound speed and damping data are also
reported from the cited INS experiment ($\blacktriangle$)
\cite{ber_ni}, as well as from MD experiments ($\circ$
\cite{alem_sim}, $-\cdot-$ \cite{ruiz}). On one hand our results
($\square$) for the acoustic excitations frequency dependence on
$Q$ are in very good agreement with the data obtained by
Ruiz-Mart\'{i}n \emph{et al.} \cite{ruiz} ($-\cdot-$) in the most
recent MD simulation on the subject, while on the other hand the
experimental INS data  ($\blacktriangle$) are qualitatively
different from both sets of data along the momentum range
investigated.\cite{ber_ni} This discrepancy is particularly
evident in the region of $Q<12\mathrm{\:nm^{-1}}$, where INS
estimated sound speed reaches the isothermal value $c_T$ (Fig.
\ref{fig:speedNi}). This could be ascribed to the kinematic
limitations to the energy accessible window at the lowest $Q$'s
which prevented the observation of the $S(Q,\omega)$ tail, and
thus a reliable estimate of the DHO model parameters.\cite{ber_ni}

\begin{figure}[ht]
\includegraphics[width=0.48\textwidth]{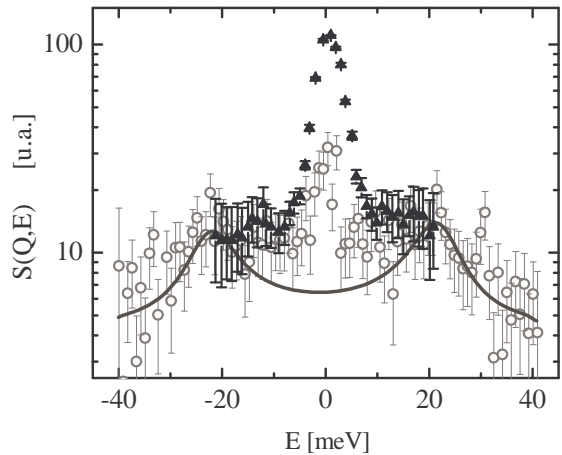}
\vspace{-6.2cm} \caption{Comparison of IXS ($\circ $) and INS
($\blacktriangle$) \cite{ber_ni} dynamic structure factors for
liquid Ni at $Q=\mathrm{8\:nm^{-1}}$. Data have been normalized to
a constant factor in order to plot them on the same scale. The DHO
function used to model IXS inelastic features is also reported
(---, see also Fig. \ref{fig:NiSpectra}). The exchanged energy
range covered by INS is narrower than that accessed by IXS, due to
the kinematics of the scattering process for neutrons.}
\label{fig:IXS_INS}
\end{figure}

In order to address the last issue, in Fig. \ref{fig:IXS_INS} we
report a comparison of IXS ($\circ $) and INS ($\blacktriangle$,
\cite{ber_ni}) dynamic structure factors at $Q=
8\mathrm{\:nm^{-1}}$, corresponding to the lowest accessed
$Q$-point for neutrons. The inelastic contribution to the IXS
spectrum is emphasized by reporting the DHO function derived from
the model (Eqs. (\ref{eq:model}) and (\ref{eq:detailedBalance})).
Beside the quasielastic peak, reflecting also the effect of
incoherent scattering (which is not present in IXS in the case of
monatomic systems),  INS data show on the anti-Stokes side the
presence of an inelastic peak around -16 meV. However, although
the detailed balance condition would imply an even more pronounced
peak on the Stokes side, its symmetric counterpart cannot be
observed.
Last but not least, the energy of propagating excitations found by
means of IXS (22.6 meV) lies just at the edge of the accessed
energy range for INS (-20 to 20 meV; see Fig. \ref{fig:IXS_INS}),
corresponding to the energy and scattering angle configurations
used in this experiment. Summing up, while at sufficiently high
$Q$'s the sound dispersion of liquid Ni lies well inside the
region of the momentum-energy range covered by INS, at $Q$ values
approaching wavevectors as low as $8\mathrm{\:nm^{-1}}$ the
reliability of the dispersion curve extracted by INS is
questionable. In Fig. \ref{fig:dampNi}.b, we report data for the
elastic linewidth, compared with the expected hydrodynamic
behavior, in which case the width of the quasielastic line is
entirely due to thermal relaxation $\Gamma_{C}=D_TQ^2$ ($---$).
Experimental INS values for $\Gamma_C$ are also shown from a
recent work ($\blacklozenge$).\cite{ruiz} The present set of data
(Fig. \ref{fig:dampNi}.b) seems to unsupport the hypothesis of a
weak Q dependence of transport coefficients.\cite{scop_rmp} Under
this circumstance, the crossover to an isothermal regime in liquid
Ni may occur at values well below $1 \div2\mathrm{\: nm^{-1}}$,
thus not accessible in the present experiment.
\\
\section{CONCLUSIONS}
In conclusion, we performed an IXS experiment on liquid Ni at 1767
K, showing the ability of such system to sustain sound propagation
over wavelengths comparable to the typical interparticle distance.
Despite the indication from a previous INS investigation
\cite{ber_ni} suggesting the tendency of sound speed to attain the
isothermal value at $Q$ lower than 8$\mathrm{\:nm^{-1}}$, we found
the evidence of an adiabatic dynamical regime holding at $Q$ as
low as 2$\mathrm{\:nm^{-1}}$, thus confirming MD results for the
microscopic dynamics of this system. \cite{ruiz, alem_sim} The
discrepancy, observed for $8<Q<12\mathrm{\: nm^{-1}}$ between INS
data on one hand and results from IXS and MD on the other hand, is
probably related to the limitations imposed to the energy-momentum
accessible region by the kinematics of the scattering process for
neutrons. Even though an adiabatic regime still holds at the
lowest accessed $Q$ for liquid Ni, the weak $Q$ dependence of the
quasielastic line-width clearly suggests that a generalized
hydrodynamic picture should be invoked. The present result calls
for further investigations with higher signal to noise ratio and
at different temperatures to unambiguously ascertain the existence
of an intermediate isothermal regime in liquid Ni in the explored
momentum region.

\end{document}